\documentclass[%
 reprint,
superscriptaddress,
 amsmath,amssymb,
 aps,
nofootinbib]{revtex4-2}

\usepackage{graphicx}
\usepackage{dcolumn}
\usepackage{color}
\usepackage[caption=false]{subfig}
\usepackage{mwe}
\usepackage{multirow}
\usepackage{amsmath}
\usepackage{bm} 
\usepackage{comment}
\usepackage[pdftex,dvipsnames]{xcolor}
\usepackage{lipsum}
\usepackage{aas_macros}
\usepackage[normalem]{ulem}

\usepackage{xargs}

\begin{document}

\preprint{APS/123-QED}

\title{
Soliton Formation and the Core-Halo Mass Relation: An Eigenstate Perspective
} 

\author{J. Luna Zagorac}
\email{lzagorac@pitp.ca}
\affiliation{
Perimeter Institute for Theoretical Physics,
31 Caroline St. N., 
Waterloo, ON N2L2Y5, Canada}
\affiliation{Department of Physics,
Yale University,
New Haven, CT 06520, USA}

\author{Emily Kendall}
\email{eken000@aucklanduni.ac.nz}
\affiliation{Department of Physics,
University of Auckland,
Private Bag 92019,
Auckland, New Zealand}

\author{Nikhil Padmanabhan}
\email{nikhil.padmanabhan@yale.edu}
\affiliation{Department of Physics,
Yale University,
New Haven, CT 06520, USA}

\author{Richard Easther}
\email{r.easther@auckland.ac.nz}
\affiliation{Department of Physics,
University of Auckland,
Private Bag 92019,
Auckland, New Zealand}

\date{\today}

\begin{abstract}
UltraLight Dark Matter (ULDM) is an axion-like dark matter candidate with an extremely small particle mass. ULDM halos consist of a spherically symmetric solitonic core and an NFW-like skirt. We simulate halo creation via soliton mergers and use these results to explore the core-halo mass relation. We calculate the eigenstates of the merged halos and use these to isolate the solitonic core and calculate its relative contribution to the halo mass. We compare this approach to using a fitting function to isolate the core and find a difference in masses up to 8.2\%. We  analyze three families of simulations---equal-mass mergers, unequal mass mergers, and halos with a two-step merger history. Setting the halo mass to the initial mass in the simulation does not yield a consistent core-halo relationship. Excluding material ``ejected'' by the collision yields a core-halo relationship with a slope of $1/3$ for simultaneous mergers and roughly 0.4  for two-step  mergers. Our findings  suggest there is no universal core-halo mass relationship for ULDM and shed light on the differing results for the core-halo relationship previously reported in the literature.
\end{abstract}

\maketitle


\newpage
\section{Introduction}

The development of accessible and accurate cosmological simulations has revolutionized the way we visualize the imprint of dark matter on observational data~\cite{2005MNRAS.364.1105S}. While simulations of  cold dark matter (CDM)  can accurately reproduce large-scale cosmological structures a number of apparent discrepancies arise on kiloparsec (kpc) scales \cite{2017ARA&A..55..343B}. One such small-scale discrepancy and the motivation for this work, the core-cusp problem, arises from differences between computationally predicted dark matter density profiles and  profiles deduced from rotation curves of dwarf galaxies \cite{2019MNRAS.486..655D, 2018MNRAS.474.1398G, 2018MNRAS.481..860R}. CDM-only simulations predict a ``cuspy'' internal profile \cite{1996ApJ...462..563N} while observations tend to favour a flatter central core \cite{1994Natur.370..629M, 2011AJ....142...24O}. Additional small-scale problems include observations of fewer satellite dwarf galaxies than simulations  suggest (the missing satellites problem) \cite{1999ApJ...522...82K, 1999ApJ...524L..19M}, an overabundance of isolated dwarfs \cite{2009ApJ...700.1779Z}, and CDM predictions of massive subhalos which are too dense not to host more bright satellites than are observed (the too big to fail problem) \cite{2011MNRAS.415L..40B}. 

The core-cusp problem is ameliorated but not necessarily solved when baryons are added  to CDM simulations \cite{2016A&A...591A..58P, 2018MNRAS.475.4825T, 2019MNRAS.488.2387B}.
Consequently, it is possible that part of this tension is due to nontrivial properties of the dark matter itself. One such example is UltraLight Dark Matter\footnote{This model is also  known as fuzzy dark matter, scalar-field dark matter, $\Psi$ dark matter, or BECDM. See Ref.~\cite{2020arXiv200503254F} for a recent review of nomenclature subtleties.} (ULDM), an axion-like particle with a mass of order $10^{-22}$ eV. The corresponding de Broglie wavelength for the ULDM particle is on kiloparsec scales, so wavelike effects can be manifest at sub-galactic scales. In particular, quantum pressure prevents the formation of a cuspy central halo, instead supporting a Bose-Einstein condensate core with a solitonic density profile~\cite{2015PhRvD..92j3513G}. Simulations demonstrate that the outer regions of ULDM halos resemble the Navarro-Frenk-White (NFW) profile of CDM \cite{1996ApJ...462..563N}, such that above kiloparsec scales the two models are mutually consistent \cite{2016PhRvD..94d3513S,  2018PhRvD..98d3509V}. A general review of ULDM physics can be found in Ref.  \cite{2017PhRvD..95d3541H}, while more recent reviews of theoretical frameworks and observational signatures can be found in Refs. \cite{2019BAAS...51c.567G, 2020arXiv200503254F}. 

Dubbed the core-halo relation, the ratio of the mass of the core to the overall halo is key to  many observational tests of ULDM. It is expected to depend on the mass of the constituent particle and the halo mass and, despite a number of analyses, disagreement persists over the specific scaling  \cite{2014PhRvL.113z1302S, Schwabe2016, 2017PhRvD..95d3519D, 2017MNRAS.471.4559M, 2020PASA...37....9K, 2022MNRAS.511..943C}. We seek to reconcile these disparate results and add a more quantitative perspective by decomposing our ULDM halos into eigenstates, via the formalism we developed in Ref.~\cite{2022PhRvD.105j3506Z}.

Our halos are formed through collisions of ULDM solitons; we arrange the solitons in  symmetric initial configurations and use the pseudo-spectral code, {\sc chplUltra} \cite{9150469}, to follow their evolution. These idealized scenarios highlight the dependence of the core-halo relationship on merger history, which will be key to mapping it to realistic astrophysical settings with hierarchical structure formation  \cite{1993MNRAS.262..627L, Masjedi:2005sc, Mackey2019}. We  find that the apparent core-halo relationship can be  impacted by numerical boundary conditions and the details of  parameter definitions. Consequently, we can contextualize many of the results seen in previous work and we see that the eigenvalue decomposition reduces scatter in the fits relative to profile-matching techniques. The overall outcome of our work suggests that we cannot find a universal core-halo relationship that is fully independent of the process by which the halo is assembled. We expect this finding to only be strengthened in more realistic halo formation scenarios that allow for effects such as mass accretion and tidal interaction with subhalos or granules. 

This paper is organized as follows: In Section \ref{sec:background}, we present the formation of halo cores through soliton collisions and discuss the dependence on initial conditions. We describe the numerical methods we use---including our three sets of data and definitions of parameters---in Section \ref{sec:numerical-methods}. In Section \ref{sec:CHMR}, we describe the scaling between halo parameters and relative core mass, and discuss the effect of merger histories. Finally, we discuss our results and future work in Section \ref{sec:discussion}.

\section{Formation of Halos}\label{sec:background}

\begin{figure*}[!ht]
    \centering
    \includegraphics[width=\textwidth]{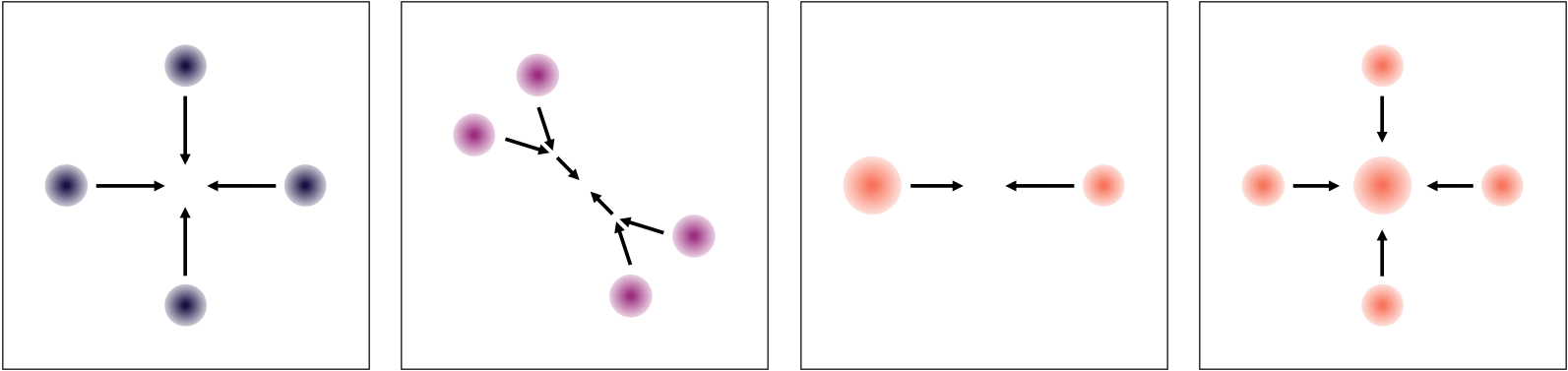}
    \caption{A schematic illustration of three data sets, left to right: simultaneous mergers, sequential mergers, and two types of unequal-mass mergers (binaries with varying mass ratio, and odd-$N$ mergers with larger central soliton.)
    }
    \label{fig:merger-schematic}
\end{figure*}

\begin{figure*}[!ht]
    \centering
    \includegraphics[width=1.05\textwidth]{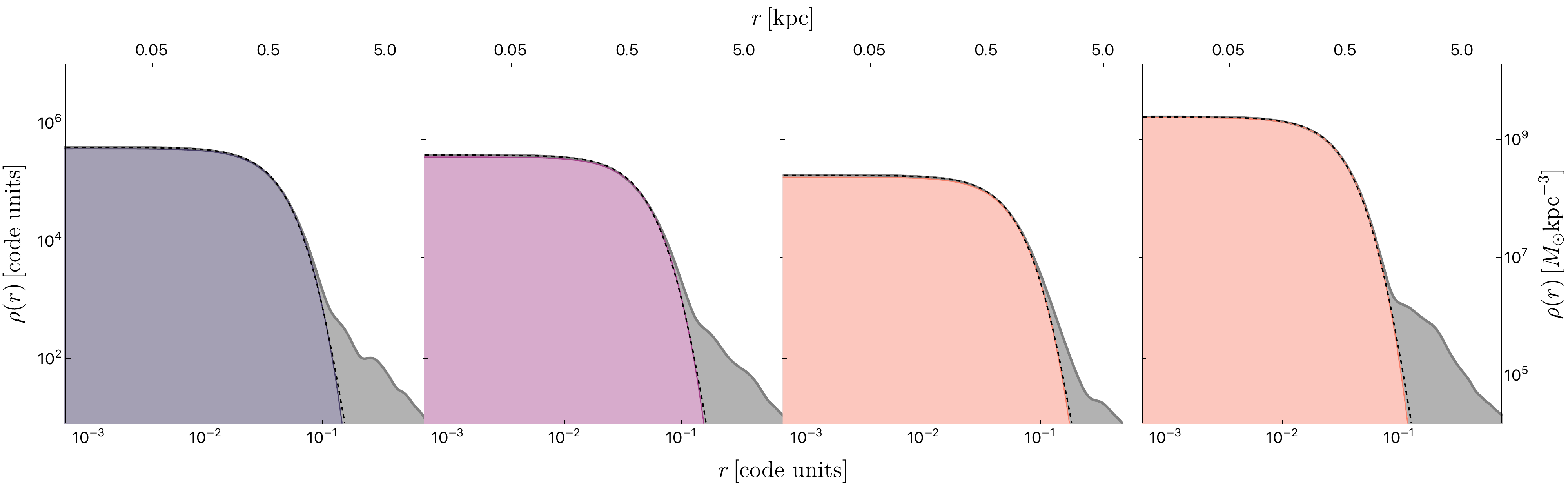}
    \caption{The halos produced through the mergers illustrated in Fig.~\ref{fig:merger-schematic}, from left to right: a simultaneous merger of four solitons, a sequential merger of four solitons, a merger of two solitons with mass ratio 1.25, and a merger of five solitons where the central soliton is 1.5 times heavier than the others. Each time- and spherically-averaged halo profile is shown in grey, with the eigenstate derived core overplotted in color. The dashed line shows the best fit of the analytic approximation of the soliton profile (Eq.~\ref{eq:solfit}). The figure is on a log-log scale. 
    }
    \label{fig:halos-and-cores}
\end{figure*}

\subsection{The Schr\"odinger-Poisson Eigensystem}

In the non-relativistic regime, ULDM can be described by a macroscopic wavefunction evolving under the influence of its own self-gravity. The dynamics are then governed by the Schr\"odinger-Poisson system of equations:
\begin{align}
i \hbar \frac{\partial \psi}{\partial t} &=-\frac{\hbar^2}{2 m} \nabla^{2} \psi+ m_a \Phi \psi \label{eq:S}\\
\nabla^{2} \Phi &=4 \pi G m |\psi|^{2}, \label{eq:P}
\end{align}
where $\psi$ is the wavefunction of the dark matter, $\Phi$ is the gravitational potential, and $m$ is the mass of the constituent particle. As the ULDM field configuration is described by a single wavefunction $\psi$, the dark matter density traces the quantum mechanical probability distribution and is defined as $\rho= m |\psi|^{2}$. The ground state of the Schr\"odinger-Poisson system is called a soliton.

Several methods of constructing generic ULDM halos can be found in the literature:  collisions of multiple solitons \cite{Schwabe2016, 2017PhRvD..95d3519D, 2017MNRAS.471.4559M, 2018PhRvD..97k6003G, 2021MNRAS.501.1539N} spherical or elliptical collapse \cite{2020PhRvD.102h3518S}, and cosmological simulations utilizing a Schr\"odinger-Poisson solver \cite{2014PhRvL.113z1302S, 2020arXiv200704119M, 2021MNRAS.506.2603M}. A fourth method \cite{2021JCAP...03..076D, 2022PhRvD.105b3512Y} uses the Widrow-Kaiser function \cite{1993ApJ...416L..71W} to build up a halo out of eigenmodes. 

We can define three pertinent initial energies in the system: the classical kinetic energy $K$, the quantum energy $Q$, and the gravitational potential energy $V$. The three components are defined as follows~\cite{2017PhRvD..95d3541H, 2017MNRAS.471.4559M}:
\begin{align}
K &= \frac{1}{2}\int d \mathbf{r} \, \rho \, v^{2} \label{eq:kinetic}\\
Q &= \frac{1}{2} \int d \mathbf{r}| \, \nabla \sqrt{\rho}|^{2} \label{eq:quantum}  \\
V &= \frac{1}{2} \int d \mathbf{r} \, \rho  \, \Phi  \label{eq:potential}
\end{align}

We create halos by coliding solitons.\footnote{Such a halo is somewhat ``synthetic", insofar as it is not produced through spherical collapse of overdensities in a cosmological code.} If the solitons are  initially non-overlapping, the kinetic term reduces to the familiar $K = \Sigma \frac{1}{2}M_i v_i^2$, where $M_i$ and $v_i$ are the mass and velocity of the $i$th soliton, respectively. The quantum energy $Q$ corresponds to the quantum pressure term which stabilizes each soliton in the absence of an internal velocity dispersion. Finally, the gravitational potential energy $V$ includes both self-gravitation and inter-soliton gravitational interactions, but is dominated by the former.\footnote{We do not consider non-gravitational self-interactions, though such models exist. For more on these, see e.g. Refs.~\cite{2016PhRvD..94h3007C, 2021PhRvD.104h3532G}.} Changing the distance between solitons thus only negligibly impacts the potential energy while changing their mass has a larger impact. For this reason, we only vary the initial masses and number of colliding solitons rather than their relative distances. Furthermore, we always start the solitons at rest. This is because we are only interested in gravitationally bound mergers where a final halo is formed, in which case adding initial velocities is physically equivalent to initializing the solitons farther away at some earlier time.

ULDM haloes consist of a solitonic core and an NFW skirt. The radial density of such a halo can be parametrized as
\begin{align}
\rho(r)=\left\{\begin{array}{ll}
\rho_{\mathrm{sol}}(r), & 0 \leq r \leq r_{\alpha} \\
\rho_{\mathrm{NFW}}(r), & r_{\alpha} \leq r \leq r_{\mathrm{vir}} \, 
\end{array}\right. \label{eq:synthetic-ULDM-halo}   
\end{align}
where $r_{\alpha}$ is the transition radius between the soliton and the skirt. The exact value of the transition radius varies in previous work (see e.g. \cite{2014PhRvL.113z1302S, 2022MNRAS.511..943C, 2022PhRvD.105b3512Y}), but is usually taken to be a few times the full width half maximum of the core, and our analysis does not depend on it at any point. The virial radius of the NFW skirt is $r_{\rm{vir}}$. Note that, while a pure soliton is an eigenstate of the Schr\"odinger-Poisson system, a composite ULDM halo is not.

In order to effectively track the evolution of the solitonic core in a halo, we first calculate the eigenstates of the Schr\"odinger-Poisson system. We begin by verifying that the potential $\Phi$ of the halo is approximately constant in time after the merger completes. This suggests the following approximation for the Schr\"odinger equation:
\begin{align}
  -i  \frac{\partial}{\partial t} \psi =
  \Biggl[-\frac{1}{2} \nabla^2 + \langle \Phi \rangle \Biggr] \psi \label{eq:Schro-avg}
\end{align}
where $\langle \Phi \rangle$ is a time-averaged potential and we have set $m_a = \hbar = G = c = 1$ for simplicity. This approximation allows us to solve for the eigenstates and eigenenergies of the system for a given $\langle \Phi \rangle$; our choices for the time-averaged potentials are discussed below. For a detailed discussion of the construction of the eigenstates, see Refs.~\cite{2022PhRvD.105j3506Z, 2021PhRvD.103b3508L}.

\subsection{Review of the Core-Halo Mass Relation}

The core-halo relationship is expressed via the unitless parameter~\cite{2016PhRvD..94d3513S,2017MNRAS.471.4559M}
\begin{equation}\label{eq:Xi}
\Xi \equiv  \frac{|E|}{ M^{3}}  \left( \frac{\hbar}{Gm} \right)^{2}   \, .
\end{equation}
Here, $E = K + Q + V$ is the total energy, and $M$ is the total mass of the system. The proposed relationship of the total halo mass and core mass $M_c$ is a power law 
\begin{align*}
M_c/M \propto \Xi^{\alpha} \,  .  
\end{align*}

Ref.~\cite{2014PhRvL.113z1302S} first suggested a relation of the form $M_c \propto (|E|/M)^{1/2}$, where the authors arrived at the $\alpha = 1/2$ result by analyzing their simulations. Their argument for the relation was as follows: the system specific energy $|E|/M$ represents the velocity dispersion of the halo $\sigma_h$, and the core mass $M_c$ is inversely proportional to the core's size, $r_c$. We can then relate these two quantities via the uncertainty principle,  $r_c \, \sigma_h \sim 1$. Finally, we can divide both sides by the halo mass $M$ to arrive at a unitless version of the relationship, which subsequent papers continued using. 

While Ref.~\cite{2019PhRvD.100l3506C} was later able to reproduce $\alpha = 1/2$ using an effective thermodynamic model, the exact scaling has been the topic of much debate. Ref.~\cite{2016PhRvD..94d3513S} found $1/6 \le  \alpha \le 1/4$ for equal-mass soliton binary mergers and $\alpha = 1/2$ for unequal-mass soliton binary mergers. Ref.~\cite{2017MNRAS.471.4559M} found $\alpha = 1/3$ in mergers of many solitons of different masses and initial conditions. Using a slightly different approach, Ref.~\cite{2017PhRvD..95d3519D} considered ULDM halos with stochastic merger trees, and found that the core-halo mass relation depends only on the mass loss fraction of cores during binary mergers, $\beta$, such that $\alpha = 2 \beta - 1$. Their data from cosmological simulations were fit well by $\beta = 0.7$, leading to an exponent of $\alpha = 0.4$. 

Refs.~\cite{2021MNRAS.501.1539N} and \cite{2020arXiv200704119M} show the results of tests using halos formed through spherical collapse in cosmological simulations; both results were consistent with $\alpha = 1/3$ rather than the original relation found in Ref.~\cite{2014PhRvL.113z1302S}. This (mis)match is somewhat curious given that Ref.~\cite{2014PhRvL.113z1302S} also performed cosmological simulations with expanding backgrounds, while Refs.~\cite{2016PhRvD..94d3513S} and \cite{2017MNRAS.471.4559M} used idealized soliton mergers to construct halos.

Ref.~\cite{2022PhRvD.105b3512Y} use a wave superposition method to artificially construct ULDM halos with a variety of density profiles. The authors then verify the dynamical stability of these halos through numerical evolution of the SP system. They are therefore able to self-consistently construct halos with different relative core sizes, further suggesting that results from simulations are not required for halo stability; rather, the cores' sizes can depend on merger history, feedback processes, etc. 

Ref.~\cite{2022MNRAS.511..943C}  discuss the range of results (including soliton mergers and cosmological simulations) and suggest that tidal stripping (both ``real" from halo interactions in the cosmological simulations and ``artificial" from choices of boundary conditions in the idealized case) contribute to the scatter in core-halo mass relations. Meanwhile, a more recent paper used approximate analytical expressions for Schr\"{o}dinger-Poisson eigenstates to show that scatter in the concentration-mass relation  contributes significantly to scatter in the core-halo mass relation, particularly for ULDM halos with mass $M > 10^9 \, \rm{M}_\odot$ \cite{2022arXiv220806562T}.

Finally, Ref.~\cite{2022arXiv220714165K} recently found that the ULDM halo profile is a better fit to Spitzer Photometry and Accurate Rotation Curves (SPARC) database than alternate models. However, the authors were unable to find a single value of particle mass that was a good fit for all the galaxies given the universal core-halo relation from Ref.~\cite{2014NatPh..10..496S}. In the case that dark matter is indeed ultralight, this result would be an argument against the existence of a universal core-halo relationship.

In this work we will primarily compare our results with references considering the same formation mechanism. Furthermore, we will consider halos of mass $M_h \lesssim 10^9 \, \rm{M}_\odot$ for our choice of units (presented in the following subsection), where Refs.~\cite{2022MNRAS.511..943C, 2022arXiv220806562T} found less scatter. In the following Sections, we will explore how the core-halo mass relation depends on merger history, as well as its sensitivity to numerical boundary conditions and show that some of these apparent discrepancies can be explained and resolved.

\section{Numerical Methods}\label{sec:numerical-methods}

Simulations in this paper are performed in  {\sc chplUltra} \cite{9150469}, a pseudo-spectral Schr\"odinger-Poisson solver. The simulation region has absorbing boundary conditions \cite{2004PhRvD..69l4033G, 2016PhRvD..94d3513S}, a grid resolution of $512^3$ and sides of length $L = 2.0$ code units, chosen to fit our largest halos while resolving our smallest cores. Each simulation is run to a time of $T=1.0$ code units, and the output is saved every $\Delta T = 0.001$. For further discussion of the numerical algorithm, see Ref. \cite{2018JCAP...10..027E}, which provides a detailed description of its implementation in a sibling code, {\sc PyUltraLight}. 

Thanks to a scaling symmetry in the Schr\"{o}dinger-Poisson equations, conversions between code units and astronomical units require only two parameters: the particle mass, $m_{22} \equiv m_a / 10^{-22}$, and a free parameter $\lambda$ (see Ref.~\cite{2018JCAP...10..027E} for details). For a convenient choice $m_{22} = 1$ and $\lambda = 2.5$, the conversion is:
\begin{align}
     \text{time code unit} &\rightarrow 12.1 \, \rm{Gyr}\\
     \text{length code unit} &\rightarrow 15.3 \, \rm{kpc} \\ 
     \text{mass code unit} &\rightarrow 5.57 \times 10^6 \, \rm{M}_\odot \, .
\end{align}
We use the above throughout this paper when converting between code units and astrophysical units. For more details on the implementation of {\sc chplUltra}, as well as more scaling choices for code units, please see Appendix A of Ref.~\cite{2022PhRvD.105j3506Z}.

\subsection{Initial conditions}

We create three separate data sets: one where $N$ equal mass solitons merge simultaneously at the center of the box, one where two groups of $N/2$ equal mass solitons merge simultaneously with other before undergoing a binary coalescence at the center of the box, and one in which solitons of unequal mass merge. Each set is schematically illustrated in Fig.~\ref{fig:merger-schematic} and described in more detail below, and four representative halos and their cores are illustrated in Fig.~\ref{fig:halos-and-cores}.

\subsubsection{Simultaneous Mergers}
    
We initialize our first set of runs with equal-mass solitons arranged in symmetric configurations, ranging from 2 to 14 solitons.\footnote{Mergers of more than 14 solitons result in final halos whose cores could not be sufficiently resolved with a $(512)^3$ grid.} We  refer to this set as our ``simultaneous mergers". The case of two solitons is a simple binary merger, while higher even numbers of solitons correspond to the vertices of the Platonic solids:  the tetrahedron (4), the octahedron (6), the cube (8), and the icosahedron (12). In addition, we set up a superposition of the octahedron and cube---called a cuboctahedron---to achieve a 14-soliton merger. Odd numbers of initial solitons are achieved by adding a single soliton to the center of the above configurations.\footnote{We do not perform a merger of 10 or 11 solitons because there is no convenient Platonic solid with 10 vertices. } We repeat simulations for three different initial soliton masses, $M = 40$, $50$ and $60$ code units and illustrate the initial conditions in the leftmost panel of Fig.~\ref{fig:merger-schematic}.  

\subsubsection{Sequential Mergers}

To initialize these merger scenarios we  populate our simulation region with $N_{\rm{sol}}$ solitons,  choosing the initial arrangement so that the mergers occur in two separate stages: two simultaneous initial mergers of $2 \geq N_{\rm{sol}} \geq 7$ solitons,  followed by a subsequent merger of the two products. Therefore, a sequential merger of e.g. 8 solitons will first involve two simultaneous 4-soliton mergers and then a subsequent binary merger. This setup is illustrated in the second panel of Fig.~\ref{fig:merger-schematic}. As before we repeat simulations for three different initial soliton masses, $M = 40$, $50$ and $60$ in code units. 

\subsubsection{Unequal mass soliton mergers}

For our final class of merger scenarios, we relax our assumption of equal progenitor masses. We do this in two batches: first, we systematically investigate the effects of changing the mass ratio $\mu \equiv M_1 / M_2$ in binary soliton mergers, where $M_1 > M_2$. Ref.~\cite{2016PhRvD..94d3513S} found that for $\mu \sim 2.5$ the less massive soliton is completely disrupted and forms a diffuse halo around the more massive progenitor, which constitutes the ``core'' of the resulting system. We therefore restrict ourselves to scenarios with $1 \leq \mu \leq 2.5$. However, we note that the total preservation of the more massive soliton as the core along with the  disruption of the less massive and thus puffier soliton does provide a mechanism for supporting a variety of relative core-halo sizes in ULDM. Specifically, we consider binary mergers with mass ratios $\mu \in \{1.05, 1.10, 1.25, 1.50, 1.75, 2.00, 2.25, 2.50\}$, with the mass of the lighter soliton being $M=40$, $M=50$, or $M=60$ in code units. These initial conditions are sketched in the third panel of Fig.~\ref{fig:merger-schematic} 

Finally, we  re-simulate each of the simultaneous mergers of odd numbers of solitons but add a central soliton with $\mu = 1.5$ times the mass of each of the other solitons, as illustrated in the last panel of Fig.~\ref{fig:merger-schematic}. In this way, we fix the mass ratio but vary the number of solitons colliding. This provides more range in $\Xi$, which is  sensitive to the number of initial solitons but  insensitive to changes in total mass. 

\begin{figure*}[ht]
    \centering
    \includegraphics[width=\textwidth]{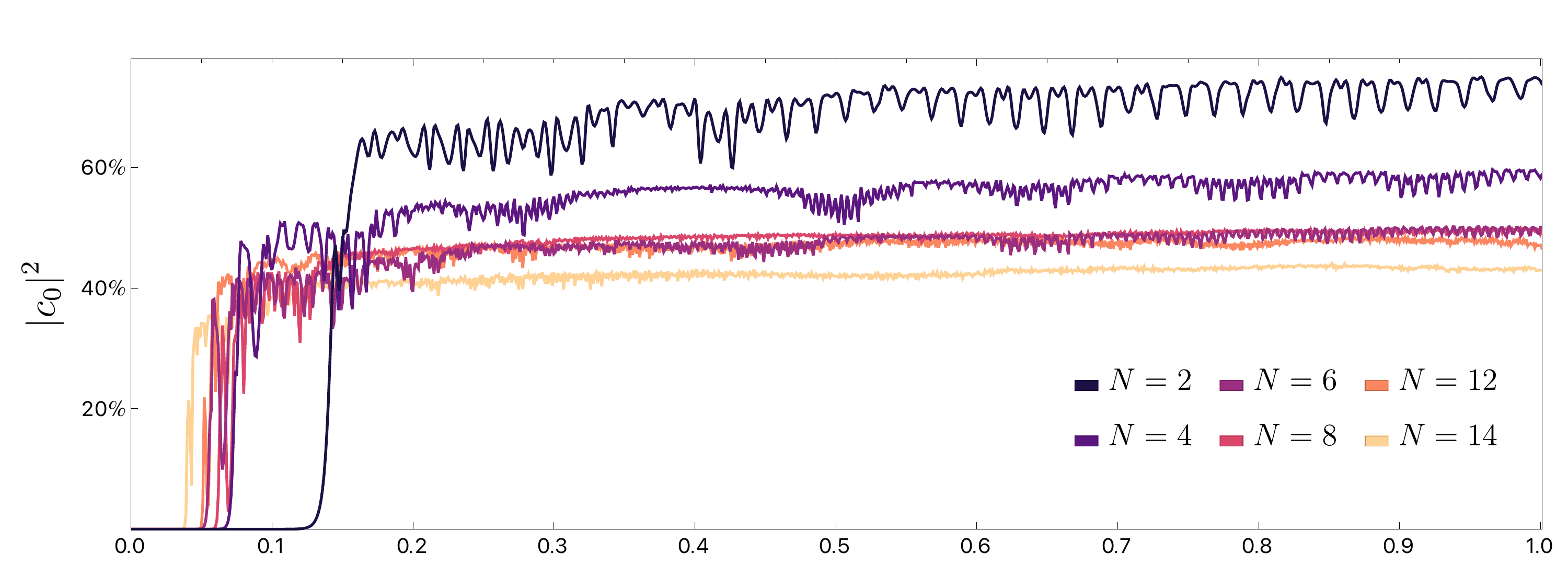}
    \caption{The evolution of the ground state of each halo is shown as a function of time. The subsample of our runs shown here are indexed by the number of simultaneously merging solitons. Note that the merger is evident from the sudden creation of a core (for instance, at $t = 0.15$ for the $N=2$ case). After the merger each halo core oscillates around an approximately constant value; we take the average from the beginning of over a period of $\Delta t = 0.2$ after each merger reaches thus quasi-static phase as the relative halo mass, $M_c/M_{h}$. For example, for the $N = 2$ merger we start averaging at time $T = 0.2$ and for $N = 14$ we begin at $T = 0.15$. Our results are insensitive to changing either the starting time $T$ or the averaging period $\Delta t$.
    }
    \label{fig:c_0}
\end{figure*}

\subsection{Relative Core Mass $M_c/M_h$}

In all of our runs the solitons start at rest, and have merged to form a halo with a core and an NFW-like skirt halfway through the simulation or earlier. We isolate the contribution of the ground state to the halo wavefunction by performing an inner product of our $\psi$-grid and the wavefunction of the ground state soliton orthonormal $\phi_0$:
\begin{align}
c_0 =\int \, d^3 r\, \psi({\mathbf r}) \phi_{0}^{*}({\mathbf r}) \,.
\end{align}
Working with the normalized wavefunction, the relative mass of the solitonic core to the halo is $|c_0|^2 = M_c / M_h$.

In order to define the soliton eigenstate $\phi_0$ in each case, we calculate the time and spherically-averaged density profile $\langle \rho \rangle$ over $0.2$ code units of time.\footnote{The halos we create should be spherically symmetric, and so a spherically-averaged density profile $\langle \rho \rangle$ is an accurate representation of the halos.} Then, we solve for the potential $\langle \Phi \rangle$ corresponding to this profile using the Poisson equation (Eq.~\ref{eq:P}), and use that the resulting potential to calculate the eigenstates of the system. We examine the ground state qualitatively to verify that it matches the shape of the core of the halo after quasi-stable behavior in $c_0$ is reached: the equivalent to a soliton of appropriate mass.  

In each case, the formation of the solitonic core is clearly seen in the sudden increase in the amplitude of $|c_0|^2$; we illustrate this effect in our simultaneous mergers data set in Fig.~\ref{fig:c_0}. After the merger, the contribution of the ground state core oscillates around an approximately constant value. Mergers with a lower mass ratio take significantly longer to ``relax"  to an approximately constant core-size after the initial merger. We start averaging only after this quasi-static phase is reached for each individual run. 

Therefore, we define the relative mass of the core, $M_c/M_{h}$, as the average of $|c_0|^2$ over $\Delta t = 0.2$ after the core starts exhibiting quasi-static behavior. We have varied both the time interval over which we average and when we begin averaging; neither impacts our results significantly. This insensitivity is important for two reasons:
\begin{enumerate}
    \item it allays any  worry that our halos have not had sufficient time to relax; and
    \item we do not need to determine exactly when the quasi-static behavior in $|c_0|^2$ begins.
\end{enumerate}
Though not shown here, the sequential and unequal mass data sets show similar behavior. Isolating the core using the eigenstates of the system is a new approach. In previous works, the mass of the core was determined by fitting the center of the halo to a density profile. In Ref. \cite{2016PhRvD..94d3513S}, the profiles are fit using the soliton profile approximation:
\begin{align}\label{eq:solfit}
\rho_{\rm{sol}}(r) \simeq \rho_{0}\left[1+0.091 \cdot\left(r / r_{c}\right)^{2}\right]^{-8} \,
\end{align}
where $\rho_0$ is the central density and $r_c$ is the FWHM of the soliton. We perform the same analysis on our data using averaged halo density profiles $\langle \rho \rangle$ over the same intervals ($\Delta t = 0.2$) as with our eigenstate analysis. We begin by setting $\rho_0 = \langle \rho \rangle (r=0)$ in the analytical profile. Next, we compute $r_c$ using the following relationship \cite{Schwabe2016}
\begin{equation}
\rho_0 \simeq 3.1 \times 10^{15}\left(\frac{2.5 \times 10^{-22} \mathrm{eV}}{m_a}\right)^2\left(\frac{\mathrm{kpc}}{r_c}\right)^4 \frac{M_{\odot}}{\mathrm{Mpc}^3}     
\end{equation}
and assuming $m_a = 10^{-22} \, \rm{eV}$. 

Once we have arrived at a soliton density profile through fitting, we calculate its mass by integrating that profile to the edge of the box. Hence, the core mass that we obtain through this procedure represents the integrated mass of the ground state
contribution over the entire halo. This approach differs from previous work wherein the soliton mass is integrated only out to a predetermined cutoff, $r_{\alpha}$ (in accordance with the halo profile of  Eq.~\ref{eq:synthetic-ULDM-halo}), as suggested in e.g. Ref.~\cite{2014PhRvL.113z1302S}. Typically $r_{\alpha}$ is a few times $r_c$, though the exact value is hard to determine precisely due to the dynamical nature of soliton cores in simulations, including the ``breathing" mode and random walk \cite{2020PhRvL.124t1301S, 2021PhRvD.103b3508L}. We argue that such an approach introduces ambiguity since the choice of $r_\alpha$ is somewhat arbitrary and computation of the core mass involves integration over higher modes in the central halo, thereby picking up non-solitonic contributions. Our eigenstate decomposition method avoids both of these issues.\footnote{Because of the steep decrease in the soliton profile outside $r_c$, our results change imperceptibly if we do impose a cutoff, further validating our approach.}

\subsection{Scaling Parameter $\Xi$}

The parameter $\Xi \equiv |E|/M^3$ can be straight-forwardly defined\footnote{We work in $G = m = \hbar = 1$ units, and so have dropped that factor from Eq.~\ref{eq:Xi}.} in a simulation where the total energy $E$ and mass $M$ in a numerical box do not vary significantly with time. However, in the course of our soliton mergers much of the halo's mass becomes unbound and ultimately removed by the absorbing boundary conditions of our box. Though the mass/energy loss decreases significantly after the merger, a small but constant mass ``leak" remains, meaning that the final value of $\Xi$ is dependent on how long we choose to run our simulation. Furthermore, the rate of this mass loss increases with rising soliton number $N$, meaning that the slope of the core-halo mass relation $M_c/M_h \propto \Xi^\alpha$ can be impacted by our choice of $\Xi$. 

With this in mind, we choose to explore two self-consistent definitions: $\Xi_i$ and $\langle \Xi \rangle$. While the final value of $\Xi$ in any run with a sponge is directly dependant on the simulation's run time, the initial value 
\begin{equation}
\Xi_i \equiv \frac{|E_i|}{M_i^3}
\end{equation}
is always robustly defined, and is equivalent to choosing to run the same simulation without a numerical sponge. The drawback of this definition is that it measures the mass and energy of the initial set of solitons arranged in the box, rather than the final halo, and so includes the contribution of mass that is no longer bound to the halo. In order to measure the latter, we use the averaged $\langle \rho \rangle$ to define
\begin{align}
\langle \Xi \rangle = \frac{|\langle K \rangle + \langle Q \rangle + \langle V \rangle|}{\langle M \rangle^3} \, .
\end{align}
The quantum energy $\langle Q \rangle$ and potential energy $\langle V \rangle$ can be directly calculated from $\langle \rho \rangle$ using Eqs.~\ref{eq:quantum} and \ref{eq:potential}. The halo mass $\langle M \rangle$ is defined as $\langle M \rangle = \int 4 \pi \langle \rho \, \rangle r^2 dr$.\footnote{Since we run idealized simulations with a non-expanding background centered on a single halo we cannot self-consistently define a critical density $\rho_c$, and therefore cannot define a meaningful virial radius.} The kinetic energy is more complex, as it requires knowledge of $\psi$ rather than $\rho$ to calculate velocities. We calculate the spherically-averaged cumulative kinetic energy out to the maximal radial distance $r = L/2$ and then average it over the same  time period as that used to fix $\langle \rho \rangle$. Given that kinetic energy is subdominant to both the quantum and potential terms, our results are insensitive to these specific choices. As with relative core mass in the previous subsection, each run is averaged over $\Delta t = 0.2$ starting when the core reaches a quasi-static state after halo-formation.
In the case where we use initial quantities, we consider $M_c / M_i \propto \Xi_i^\alpha$ and in the averaged case we consider $M_c / \langle M \rangle \propto \langle \Xi \rangle^\alpha$. In both cases we do not define the halo mass in terms of a virial radius, but integrate over the whole box (or, in the case of $\langle M \rangle$, over the largest sphere inscribed in our box).

\section{Core-Halo Mass Relation(s)}\label{sec:CHMR}

\subsection{Simultaneous Mergers}

\begin{figure}[!ht]
    \centering
    \includegraphics[width=\columnwidth]{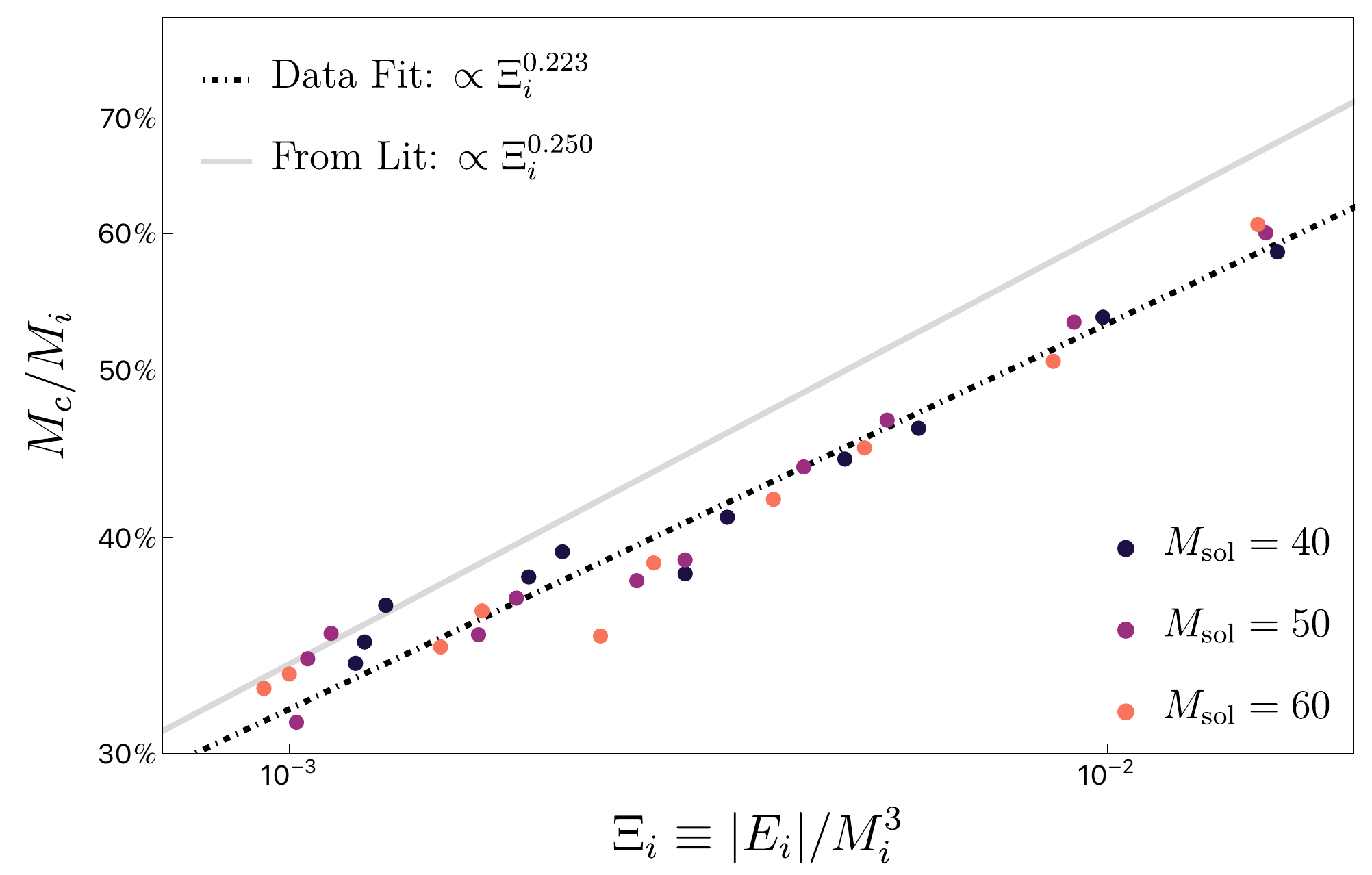}
    \includegraphics[width=\columnwidth]{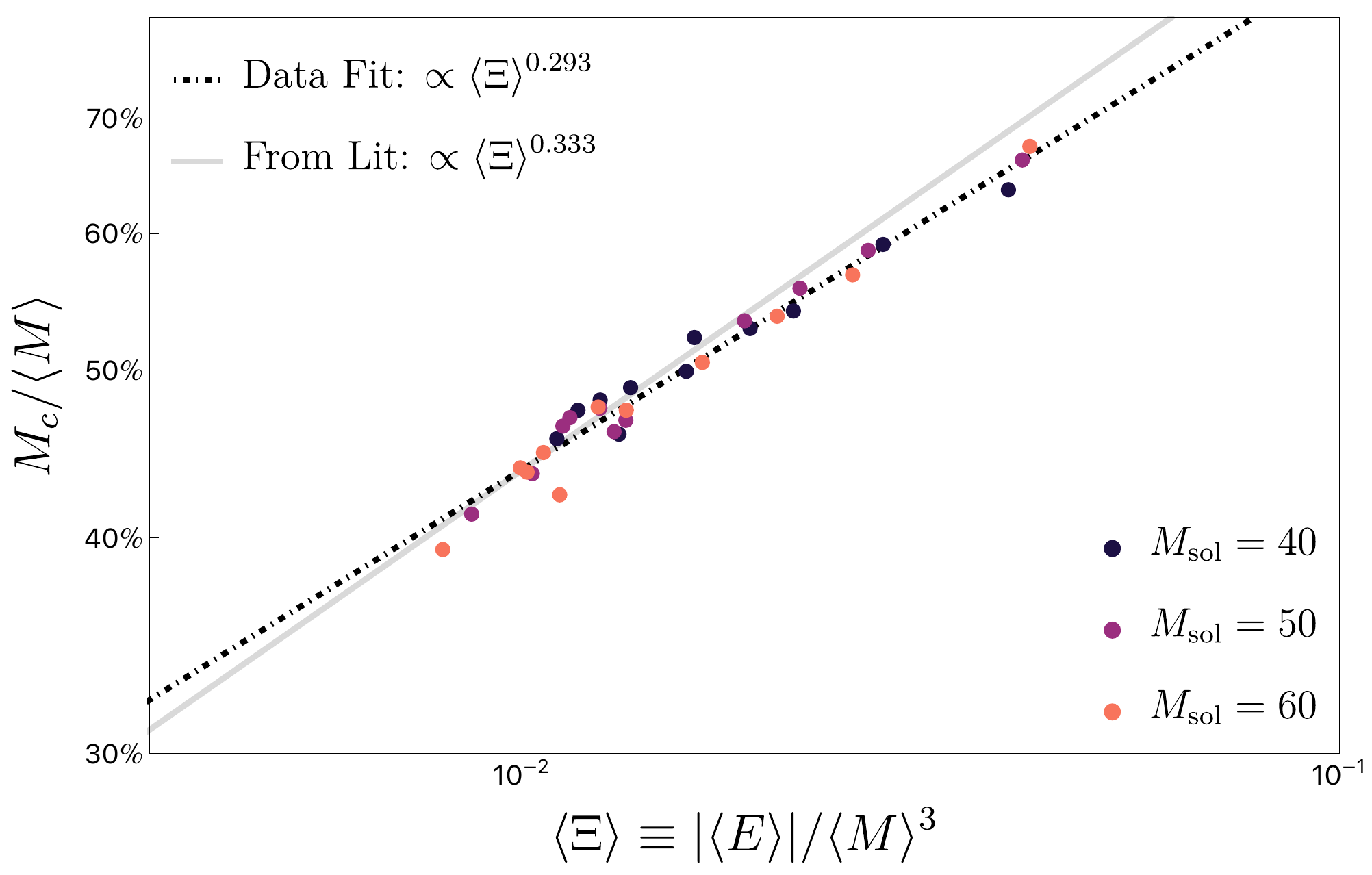}
    \caption{We present the log-scaled relationship between the relative core mass and our two $\Xi$ parameters in our simultaneous data. The data points are colored by the initial mass of the merging solitons, $M_{\rm{sol}}$, in code units, and their best fit slope is shown by the black dot-dashed line. Both panels show the same range in $\Xi$, shifted by a factor of 5.\\
    \textbf{Top:} Using $\Xi_i$, the power law fit to our data yields an exponent of $0.223$ with a 95\% confidence interval of $0.209 - 0.237$. A line of slope $1/4$ is shown in gray for comparison with the closest result from previous literature (Ref.~\cite{Schwabe2016}).\\
    \textbf{Bottom:} Using $\langle \Xi \rangle$, the power law fit to our data yields an exponent of $0.293$ with a 95\% confidence interval of $0.277 - 0.308$. A line of slope $1/3$ is shown in gray for comparison with the closest result from previous literature (Ref.~\cite{Mocz2017}).
    }
    \label{fig:CHMR}
\end{figure}

We calculate the parameter $\Xi$ using our two definitions and plot it against our derived relative core masses showing the results in Fig.~\ref{fig:CHMR}. Fitting a power law to our data, we find a slope $\alpha = 0.223$ when using $\Xi_i$ and $\alpha = 0.293$ when using $\langle \Xi \rangle$. Both plots are shown on the same vertical axis and with the same range on the horizontal axis, but shifted relative to one another. In both panels the data points are roughly groups in threes, corresponding to the three initial soliton masses: $M_{\rm{sol}} = 40, 50$ and $60$ (in code units). 

In both plots there is more visible scatter  as $\Xi \rightarrow 0$, corresponding to more massive halos formed through collisions of higher numbers of solitons. For $\Xi_i$, this stems partially form the fact that the more solitons we collide, the less the initial conditions in the box describe the final halo produced through the collision.  $\langle \Xi \rangle$, on the other hand, does describe the actual halos; however, the range of $\langle \Xi \rangle$ changes due to our accounting for the mass and energy loss in the box. Consequently, the confidence intervals returned by the fitting function are comparable for both choices of $\Xi$. 

\begin{figure}[htpb]
    \centering
    \includegraphics[width=\columnwidth]{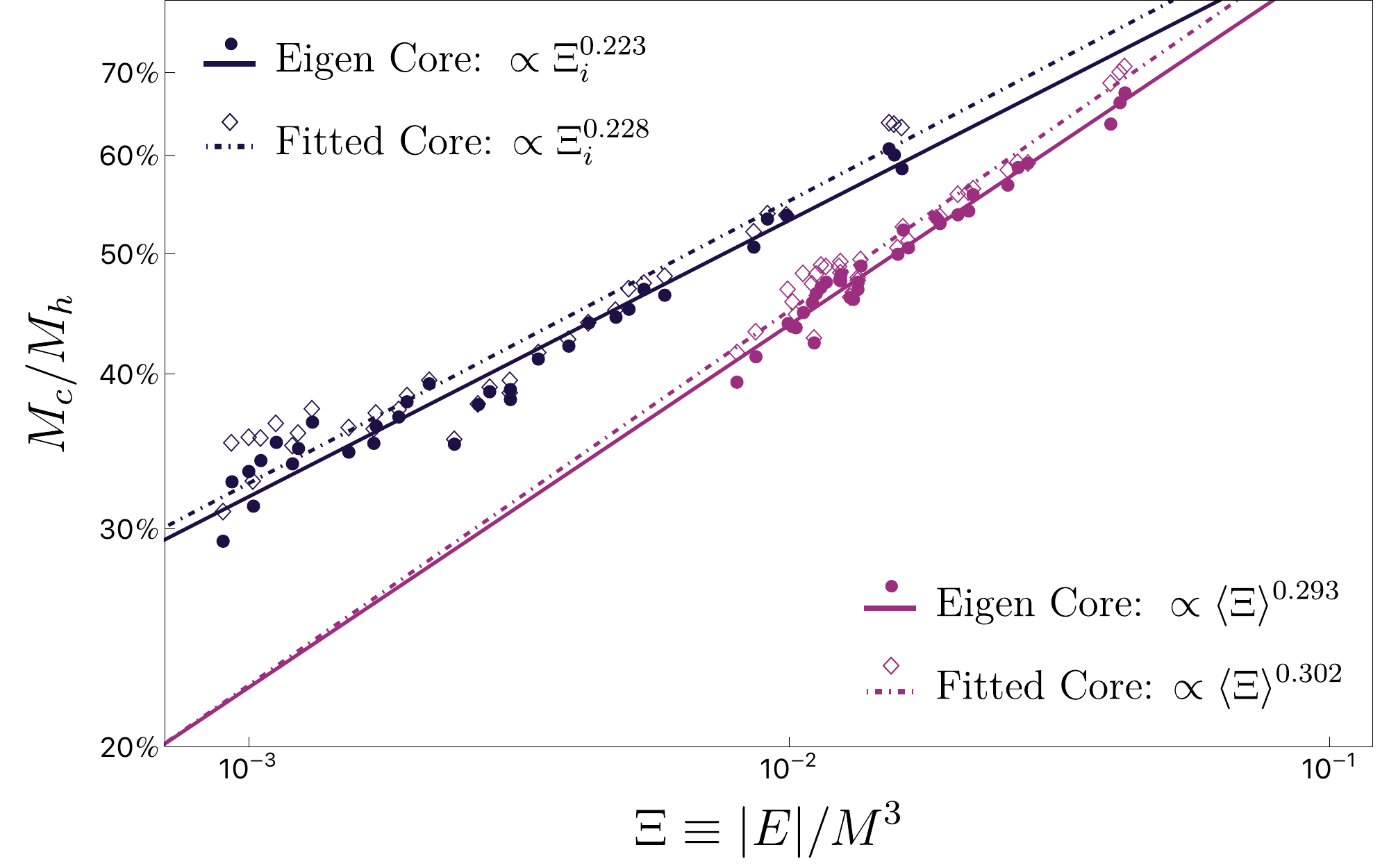}
    \caption{Shown is the core-halo mass relation for cores derived from eigenstates (full circles, full lines) and from fitting functions (hollow diamonds, dot-dashed lines). The points show individual halos, while the lines are the best fit slopes indicated in the legends. The black data use $\Xi_i$, the pink data use $\langle \Xi \rangle$. All sets of data include mergers of $M_{\rm{sol}}=40$, $50$, and $60$ solitons. 
    }
    \label{fig:CHMR-fitVSeig}
\end{figure}

We illustrate the results from fitting cores and compare them with results from calculating soliton eigenstates in Figure~\ref{fig:CHMR-fitVSeig}. Again, the two choices of $\Xi$ produce different slopes, and the fitted cores underestimate the slope (and relative core size) when compared to eigenstate cores for both definitions. In each case, the eigenstate-derived cores exhibit slopes closer to numbers previously put forth in the literature: $1/5$ and $1/3$ for $\Xi_i$ and $\langle \Xi \rangle$, respectively. Comparing eigenstate and fit cores directly, we find a difference of up to 8.2\%, with the fitting method mostly overestimating core mass compared to the eigenstate equivalent. Nevertheless, the best fit slope for each choice of $\Xi$ is not dissimilar to the equivalent eigenstate-extracted cores. 

\begin{figure}[htpb]
    \centering
    \includegraphics[width=\columnwidth]{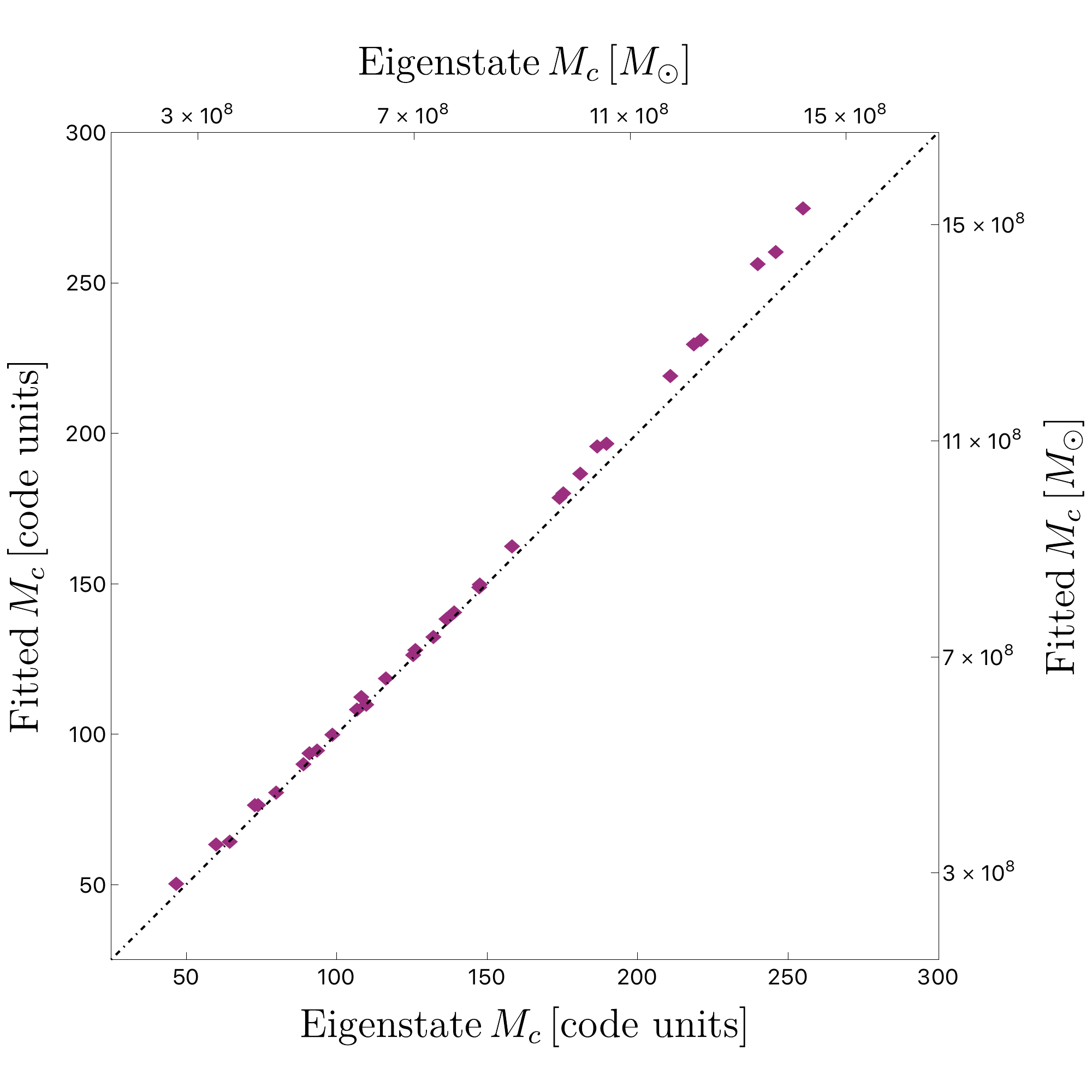}
    \includegraphics[width=\columnwidth]{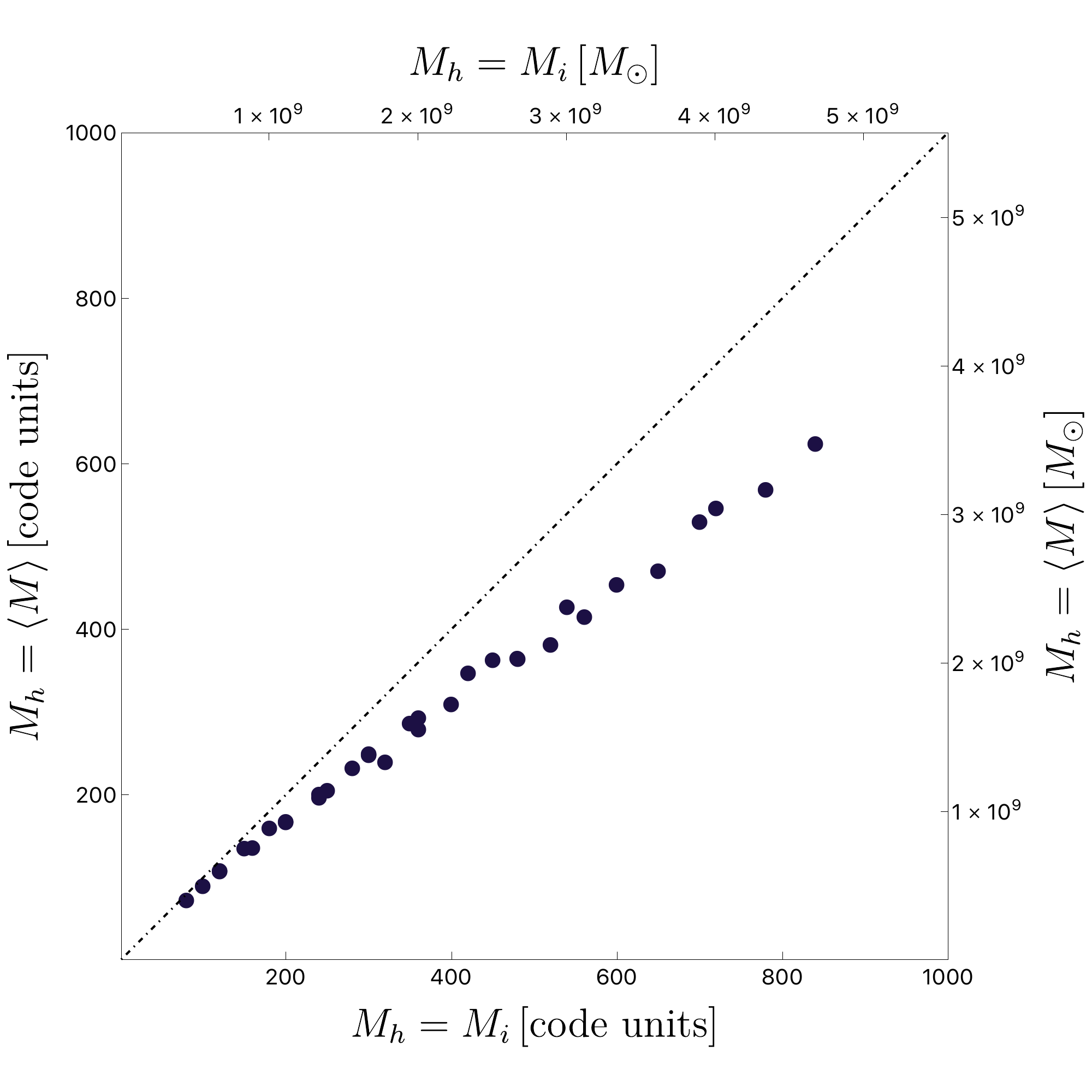}
    \caption{We show the difference between our two core estimation methods (eigenstates and fitting), as well as the difference between the two halo mass definitions, $M_i$ and $\langle M \rangle$. The dot-dashed line in both panels represents a slope of one as an aid to the reader, and masses are presented both in code units and solar masses $\rm{M}_\odot$.
    }
    \label{fig:M_vs_M}
\end{figure}

Furthermore, the 95\% parameter confidence interval on the fitted $\Xi_i$ data is $0.209 - 0.246$; for $\langle \Xi \rangle$ it is $0.281 - 0.323$. These are comparable to the confidence intervals quoted in Fig.~\ref{fig:CHMR}, suggesting that while the two core extraction approaches perform similarly well. This is to say: while we prefer our eigenstate method of defining the core as more theoretically robust and unambiguous, it does not significantly lessen the scatter in our data. Therefore, the information presented in this subsection suggests that---for a single formation mechanism, at least---the definition of the halo mass $M_h$ has a much stronger effect on the core-halo mass relation than the definition of the halo core mass, $M_c$. We verify this in Fig.~\ref{fig:M_vs_M}: while the choice of core definition matters very little for our halos, the choice of halo mass ($M_i$ or $\langle M \rangle$) is quite impactful for all but the lowest mass halos. 

\subsection{All Data}

\begin{figure}[ht]
    \centering
    \includegraphics[width=\columnwidth]{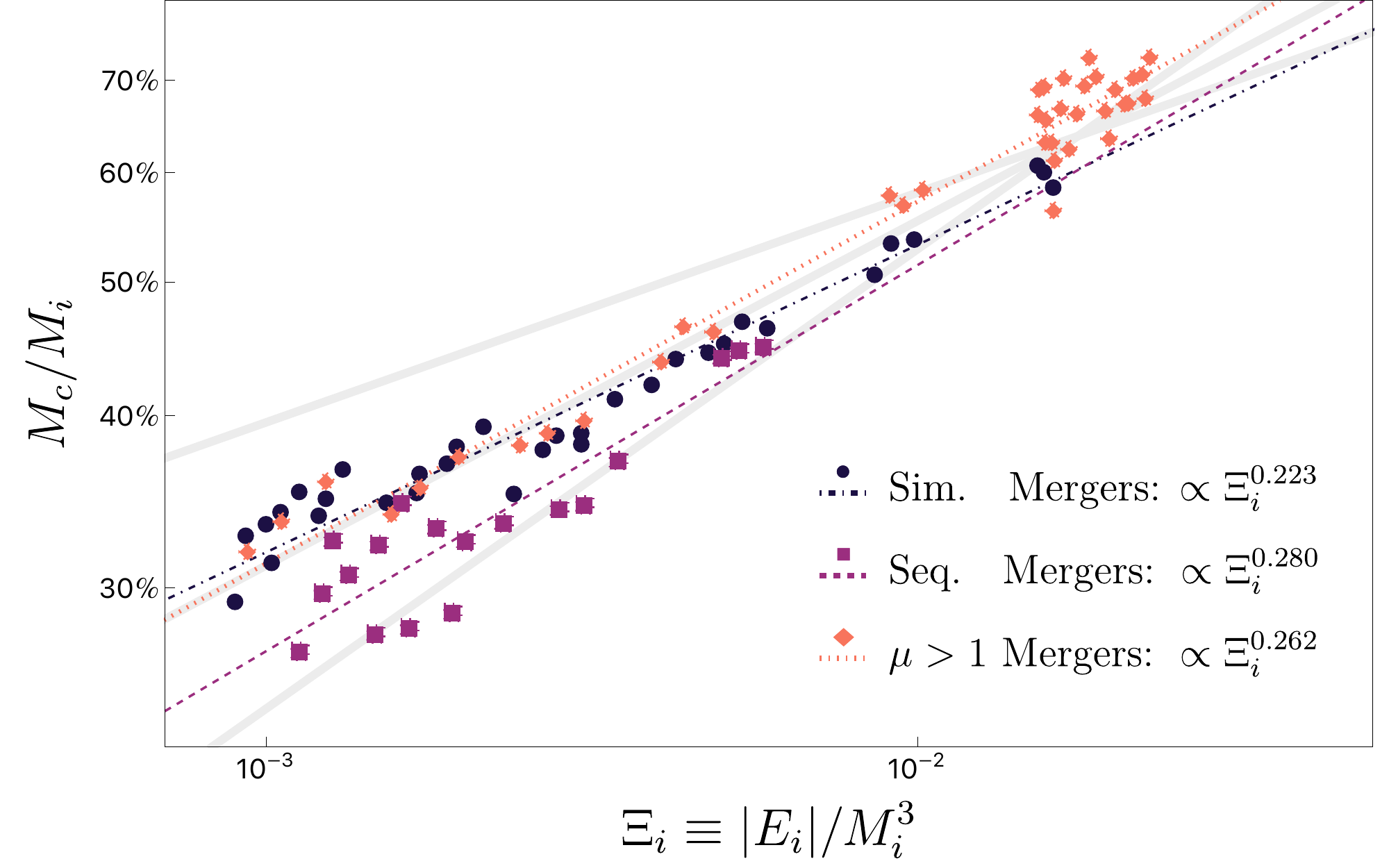}
    \includegraphics[width=\columnwidth]{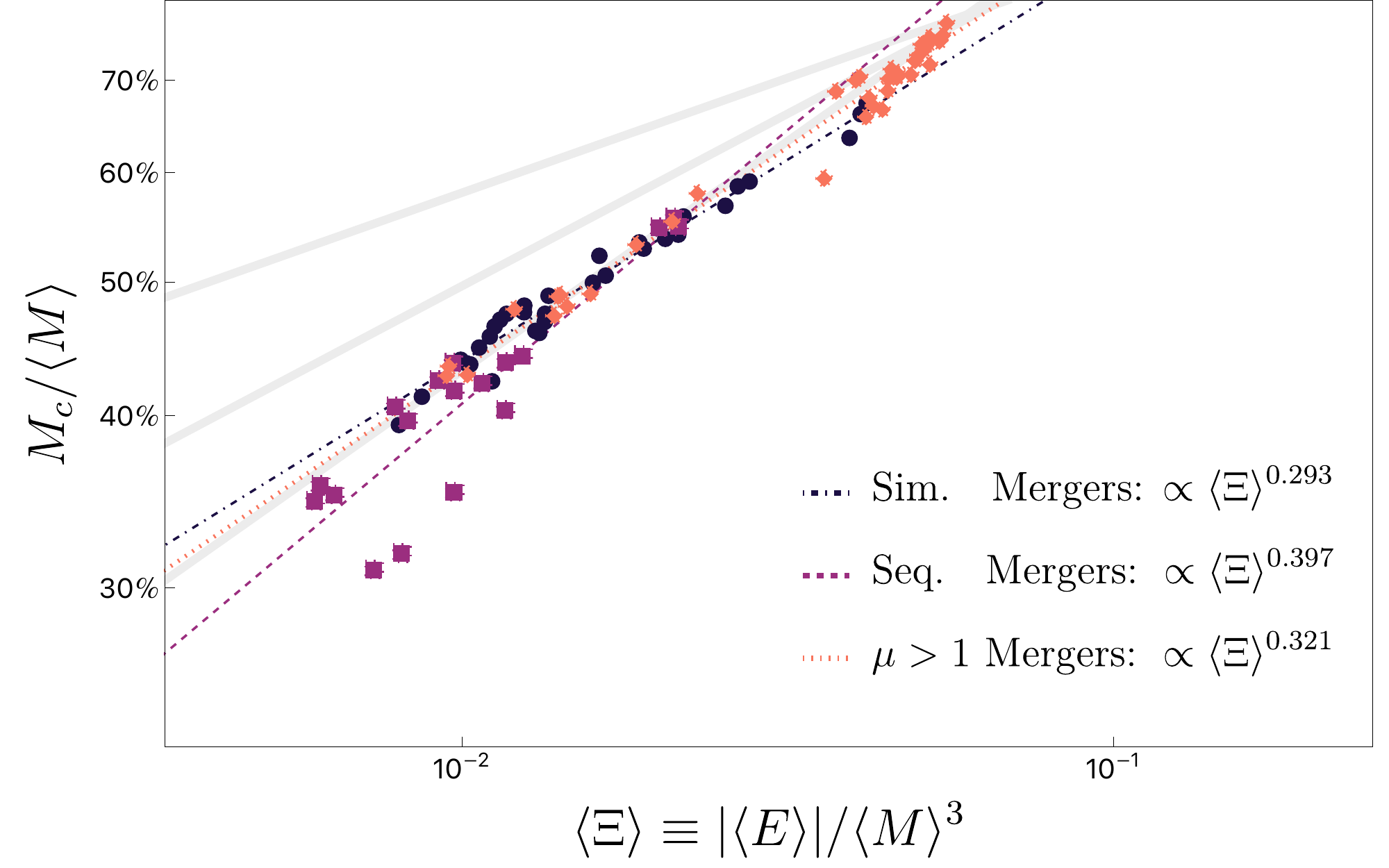}
    \caption{We present the log-scaled relationship between the relative core mass and our two $\Xi$ parameters in all three data sets. The grey lines show exponents of $1/3$, $1/4$, and $1/6$ for comparison. Both panels show the same range in $\Xi$, shifted by a factor of 5. In both panels, the simultaneous data (best fit slope) are represented by black circles (dot-dashed line), the sequential data by purple squares (dashed line), and the mass ratio data by orange diamonds (dotted line). \\ 
    \textbf{Top:} Using $\Xi_i$, the power law fit to our data yields exponents (95\% confidence intervals) of $0.223 \, (0.209 - 0.237)$, $0.280 \, (0.219 - 0.341)$, and $0.262 \, (0.265 - 0.292)$. \\
    \textbf{Bottom:} Using $\langle \Xi \rangle$, the power law fit to our data yields an exponents (95\% confidence intervals) of $0.293 \, (0.277 - 0.308)$, $0.397 \, (.313 - 0.481)$, and $0.321 \, (0.303 - 0.339)$. 
    }
    \label{fig:CHMR-all}
\end{figure}

We now include our sequential and unequal mass data sets in the analysis; the results are shown in Fig.~\ref{fig:CHMR-all}. In the case of $\Xi_i$, each data set produces a different best slope: $0.223$ (simultaneous), $0.280$ (sequential), and $0.262$ (unequal mass). There is significant  scatter around each of the fits, particularly towards  low-$\Xi$ (high mass) values and in the cluster of unequal mass binary mergers   in the upper right corner. There is no clear linear relationship in the data.

For $\langle \Xi \rangle$, the best fit slopes are more uniform:  $0.293$ (simultaneous), $0.397$ (sequential), and $0.321$ (unequal mass). For the simultaneous and sequential mergers the scatter in visually smaller, and both data can be approximately described with a single line of slope $\alpha \approx 1/3$. The sequential mergers favor a slope $\alpha \approx 0.4$ (although $\alpha = 1/3$ falls within the fit's 95\% confidence interval) and show more scatter as $\langle \Xi \rangle \rightarrow 0$.

The $\Xi_i$ values presented in the upper panel of Fig.~\ref{fig:CHMR-all} are consistent with the findings of Ref.~\cite{2016PhRvD..94d3513S} and a slope of approximately $1/4$. The authors of Ref.~\cite{2016PhRvD..94d3513S} simulated binary mergers of solitonic cores in the presence of a sponge at the numerical boundary. Furthermore, they normalized their core masses with respect to the initial mass of the simulation; thus, their setup is most similar to our simultaneous and unequal mass binary mergers analyzed against $\Xi_i$. Given the scatter in our data and its evident dependence on merger history we are not able to claim a single well-defined core-halo mass relation. 

The $\langle \Xi \rangle$ data can be fit by a single line with a slope of approximately $1/3$, though sequential mergers diverge both in terms of best-fit slope (0.4) and increased scatter. The result of sequential mergers matches the findings of Ref.~\cite{2017PhRvD..95d3519D}, in which the authors explicitly considered merger history. The 1/3 result matches the analysis of Ref.~\cite{2017MNRAS.471.4559M} which investigated scenarios involving a group of solitonic cores of different masses that merge sequentially to form a final virialized halo in a numerical box without sponge boundary conditions. They worked with  $\Xi_i$  and  we reproduce their results \textit{despite} our differing numerical setup and definition of $\Xi$, suggesting that merger history plays a stronger role in the final core-halo relationship than the exact definition of $\Xi$. We thus also recover the argument of Ref.~\cite{2022MNRAS.511..943C}, where the authors suggest that a diversity of core-halo mass relations can partially be attributed to ``artificial" stripping due to boundary condition choices. However, we are able to somewhat blunt this effect by using the definition $\langle \Xi \rangle$. 

We attribute this to the robustness of the $\langle \Xi \rangle$ definition, stemming directly from the definition of $\langle \rho \rangle$. By employing a spherically- and time-averaged definition of our halo profile after the core has reached the steady state, we construct a definition of halo mass and energy that varies less with time. Additionally, by using $\langle \Xi \rangle$ we average over spurious contributions to the edge of the halo arising from the numerical boundary. Finally, this definition is the more useful one when considering real cosmological halos, as we will not have access to information on the initial masses and energies of individual halos. 
However, it's worth noting that using instantaneous values for astrophysical halos would introduce scatter into the observed core-halo relationship (as in Ref.~\cite{2020PASA...37....9K}) when compared the simulated one using averaged quantities; nevertheless, we expect the main result to stay the same. 

\begin{figure}[ht]
    \centering
    \includegraphics[width=\columnwidth]{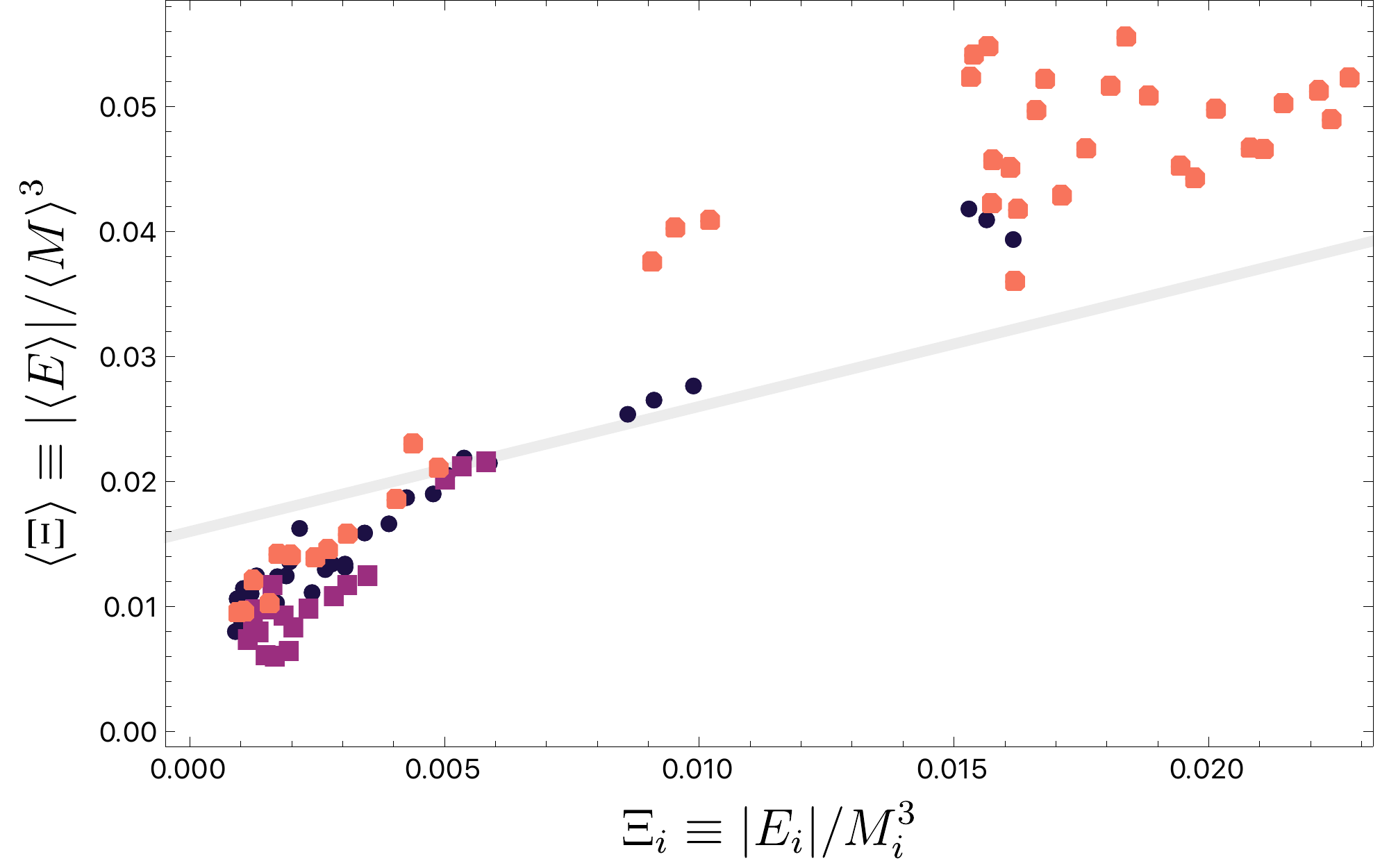}
    \caption{We show the relationship between our two $\Xi$ definitions, $\Xi_i$ and $\langle \Xi \rangle$, for all our available data. The data are plotted with the same color and shape scheme as in Fig.~\ref{fig:CHMR-all}, and the gray line denotes a slope of 1.   
    }
    \label{fig:Xi_vs_Xi}
\end{figure}

Finally, we compare our two definitions of $\Xi$ directly in Fig.~\ref{fig:Xi_vs_Xi}. We note that while the relationship between $\Xi_i$ and $\langle \Xi \rangle$ is approximately the same for our simultaneous and different mass data sets, the sequential data once again have a distinct slope. This provides further evidence to our conclusion that we cannot expect a single power law in $\Xi$ to describe populations of halos with different merger histories. 

\section{Discussion}\label{sec:discussion}

We have presented a novel approach to isolating the cores of ultralight dark matter halos by calculating the ground state of the central soliton. We create halos through mergers of different numbers  of solitons of varying masses. The ground state (and, indeed, the higher states) is calculated using the  approximately stationary potential of the isolated halo, as described in Ref.~\cite{2022PhRvD.105j3506Z}. We  then project the halo wavefunction onto the basis of eigenstates at each timestep, revealing the time evolution of the halo core. Following the initial  formation phase the contribution of the ground state is approximately constant, allowing us to define a relative core-halo mass, $M_c/M_{h}$.

This approach  illuminates earlier approaches to the problem. Previous work (starting with Refs.~\cite{2014PhRvL.113z1302S, 2014NatPh..10..496S}) has established the core-halo  relation as a function of $M_c/M_{h}$.  This is usually taken to be a power law $M_c/M_{h} \propto \Xi^\alpha$, where $\Xi \equiv |E|/M^3$ and values of $\alpha$ ranging from $1/6$ to $1/2$ have been suggested in the literature. Here, we explore  both $\Xi_i$, calculated with respect to the initial mass $M_i$ and energy $E_i$ in the box, and $\langle \Xi \rangle$, calculated from the spherically- and time-averaged halo density profile $\langle \rho \rangle$.

We compared fits to an analytic core+NFW profile and the eigenstate approach, and fitting slightly overestimates the slope of the core-halo relationship for both definitions of $\Xi$ relative to the eigenvalue method, such that $\alpha_{\rm{fit}} > \alpha_{\rm{eig}}$. The core masses differ up to 8.2\%, with fitted cores being more massive than their eigenstate-derived counterparts. These discrepancies are relatively small and may follow from fitting to the density of the halo rather than its wavefunction:  the former is the square of the latter, $\rho = |\psi|^2$, so  the density may include cross-terms which are not present when dealing with the wavefunction directly. In any case, the similarity between the two methods  suggests that the scatter in the core-halo mass relation is due to processes affecting the halo skirt rather than the core. 

We simulated soliton collisions in three regimes:
\begin{enumerate}
    \item simultaneous mergers---where  $N$ solitons of identical masses merge collide at the center of our box;
    \item sequential mergers---where two groups of $N$ same-mass solitons merge at the same time, after which the two  merger products collide at the center of the box, forming the final halo; and
    \item unequal mass mergers---mergers of two solitons with a mass ratio, $\mu = M_1/M_2$, or merging odd-$N$ with the central soliton having a mass 1.5 times larger than the others.
\end{enumerate} 

Working with  $\Xi_i$, we do not find a single scaling behavior between the three data sets, and even  within individual data sets there is substantial scatter. The slopes range from $\alpha \approx 1/5$ to $\alpha \approx 1/3$. The latter is consistent with the findings of Ref.~\cite{2016PhRvD..94d3513S} who simulated binary mergers of solitonic cores in the presence of a sponge at the numerical boundary (as we do) and used the $\Xi_i$ definition. For  $\langle \Xi \rangle$, the overall best fit is $\alpha \approx 1/3$, though the sequential mergers are better fit with $\alpha \approx 0.4$ and display more scatter. The 1/3 slope matches the results of Refs.~\cite{2017MNRAS.471.4559M, 2021MNRAS.501.1539N}, while the 0.4 slope matches those of Ref.~\cite{2017PhRvD..95d3519D}. 

The scatter and different slope in the sequential data suggest that merger history plays a non-negligible part in determining the relative mass of halo cores, possibly introducing a non-trivial redshift dependence (e.g. as in  Ref.~\cite{2022arXiv220806562T}). Furthermore, the simulations here are idealised scenarios and astrophysical environments will be more complex. In particular, cosmological data gives us instantaneous snapshots of galaxies rather than time-averaged quantities, which will introduce additional scatter to any observationally motivated core-halo relationship \cite{2020PASA...37....9K}. Thus, our analysis does not support the existence of a single and
universal core-halo mass relation for ULDM halos.  Moreover, the inclusion of more complicated physics in halos such as mass accretion, dynamical heating from granules, or tidal interactions with subhalos,  may further impact the core-halo relationship.

Finally, the underlying dynamics of ULDM---self-gravitating quantum matter that undergoes gravitational collapse---may be replicated in the post-inflationary universe \cite{Musoke:2019ima,2020JCAP...07..030N, 2021PhRvD.103f3525E,2022PhRvD.105b3516E,2022arXiv221200425E,2022arXiv221200425E} and in hypothetical axion miniclusters \cite{Hogan:1988mp,Ellis:2022grh}. Soliton-like structures can form in both these phases and could play a role in a long post-inflationary matter-dominated phase, and these will like obey a similar core-halo relationship to ULDM. Consequently, the analysis here can shed light on these scenarios, which are governed by a common dynamical system, despite their very different physical basis.

\begin{acknowledgments}
  We thank the Cray/HPE Chapel team, especially Elliot Ronaghan, for collaborating on the development
  of {\sc chplUltra} and for the computational resources used in this paper. While at Yale, JZ was supported by the Future Investigations in NASA Earth and Space Science and Technologies (FINESST) grant (award number 80NSSC20K1538). Research at Perimeter Institute is supported in part by the Government of Canada through the Department of Innovation, Science and Economic Development Canada and by the Province of Ontario through the Ministry of Colleges and Universities. RE and EK acknowledge support from the Marsden Fund of the Royal Society of New Zealand. 
\end{acknowledgments}



\bibliographystyle{apsrev4-2}
\bibliography{bibliography.bib}

\end{document}